\documentclass[twoside,12pt]{article}

\usepackage[british]{babel}
\usepackage[T1]{fontenc}

\usepackage{cite}                         
\usepackage{citesort}                     
\usepackage[dvips]{color}                 
\usepackage{graphicx}                     
\usepackage{amssymb}
\usepackage{amsmath}
\usepackage{xspace}                       

\newcommand{\absatz}{\vspace{2ex}\noindent}

\newcommand{\blue}[1]{#1}
\newcommand{\red}[1]{#1}

\newcommand{\green}[1]{#1}

\newcommand{\EPJA}{\emph{Eur.\  Phys.\ J.\ } {A}}

\def\NPA{\emph{Nucl. Phys.} A}

\def\PLB{\emph{Phys. Lett.} B}

\def\PLB{\emph{Phys. Lett.} B}
\def\PRL{\emph{Phys. Rev. Lett.}}

\def\PRC{\emph{Phys. Rev.} C}

\def\ANNP\emph{Ann. Phys. (N.Y.)}

\newcommand{\dis}{\displaystyle}

\newcommand{\non}{\nonumber}

\newcommand{\half}{\frac{1}{2}}

\newcommand{\ii}{\mathrm{i}}

\newcommand{\MeV}{\ensuremath{\mathrm{MeV}}}
\newcommand{\fm}{\ensuremath{\mathrm{fm}}}
\newcommand{\ChiEFT}{$\chi$EFT\xspace}

\newcommand{\de}{\partial}

 \newcommand{\calO}{\mathcal{O}}

\begin{document}


\title{\vspace*{1cm}Nucleon Polarisabilities from Compton Scattering\\
  off the One- and Few-Nucleon System}
              
\author{Harald W.\ Grie\3hammer${}^{1}$\footnote{Email:
    hgrie@physik.tu-muenchen.de. Supported by DFG under contract GR 1887/2-2.
}~\footnote{Preprint TUM-T39-04-16,
                  nucl-th/0411080. Invited seminar given at the \textsc{26th Course of
                    the International School of Nuclear Physics: Lepton
                    Scattering and the Structure of Hadrons and Nuclei}, Erice
                  (Italy), 16th -- 24th September 2004. To be published in
                  \emph{Prog.~Nucl.~Part.~Phys.} \textbf{54}, No.~2 as part of
                  the proceedings.
}\\ \\ 
$^1$Institut f{\"u}r Theoretische Physik (T39), Physik-Department,\\
Technische Universit{\"a}t M{\"u}nchen, D-85747 Garching, Germany} \maketitle
\begin{abstract} \noindent
  These proceedings sketch how combining recent theoretical advances with data
  from the new generation of high-precision Compton scattering experiments on
  both the proton and few-nucleon systems offers fresh, detailed insight into
  the Physics of the nucleon polarisabilities. A multipole-analysis is
  presented to simplify their interpretation. Predictions from Chiral
  Effective Field Theory with special emphasis on the spin-polarisabilities
  can serve as guideline for doubly-polarised experiments below $300$ MeV. The
  strong energy-dependence of the scalar magnetic dipole-polarisability
  $\beta_{M1}$ turns out to be crucial to understand the proton and deuteron
  data. Finally, a high-accuracy determination of the proton and neutron
  polarisabilities shows that they are identical within error-bars. For
  details and a better list of references, consult
  Refs.~\cite{Griesshammer:2001uw,polas2,polas3,dpolas}.
\end{abstract}

%
%

\section{Introduction}
\label{sec:introduction}

Nuclear physicists are hardly surprised by the fact that the nucleon is not a
point-like spin-$\half$ target with an anomalous magnetic moment in low-energy
Compton scattering $\gamma N\to\gamma N$. Rather, the photon field displaces
its charged constituents, inducing a non-vanishing multipole-moment. These
nucleon-structure effects have in fact been known for many decades and (in the
case of a proton target) quite reliable theoretical calculations for the
deviations from the Powell cross-section exist. They are canonically
parameterised starting from the most general interaction between the nucleon
$N$ with spin $\vec{\sigma}/2$ and an electro-magnetic field of fixed,
non-zero energy $\omega$:
\begin{equation}
\begin{split}
  \label{polsfromints}
  &2\pi\;N^\dagger \;\big[\red{\alpha_{E1}(\omega)}\;\vec{E}^2\;+
  \;\red{\beta_{M1}(\omega)}\;\vec{B}^2\; +\;\red{\gamma_{E1E1}(\omega)}
  \;\vec{\sigma}\cdot(\vec{E}\times\dot{\vec{E}})\;
  +\;\red{\gamma_{M1M1}(\omega)}
  \;\vec{\sigma}\cdot(\vec{B}\times\dot{\vec{B}}) \\&\;\;\;\;\;\;\;\;\;\;\;\;
  -\;2\red{\gamma_{M1E2}(\omega)}\;\sigma_i\;B_j\;E_{ij}\;+
  \;2\red{\gamma_{E1M2}(\omega)}\;\sigma_i\;E_j\;B_{ij} \;+\;\dots \big]\;N
\end{split}
\end{equation} 
Here, the electric or magnetic ($\blue{X,Y=E,M}$) photon undergoes a
transition $\blue{Xl\to Yl^\prime}$ of definite multipolarity
$\blue{l,l^\prime=l\pm\{0,1\}}$; $\green{T_{ij}:=\half (\de_iT_j +
  \de_jT_i)}$, and the coefficients are the \emph{energy-dependent} or
\emph{dynamical polarisabilities} of the nucleon.  Most prominently, there are
six dipole-polarisabilities: two spin-independent ones ($\alpha_{E1}(\omega)$,
$\beta_{M1}(\omega)$) for electric and magnetic dipole-transitions which do
not couple to the nucleon-spin; and in the spin-sector, two diagonal
(``pure'') spin-polarisabilities ($\gamma_{E1E1}(\omega)$,
$\gamma_{M1M1}(\omega)$) and two off-diagonal (``mixed'')
spin-polarisabilities, $\gamma_{E1M2}(\omega)$ and $\gamma_{M1E2}(\omega)$. In
addition, there are negligible contributions from higher ones like quadrupole
polarisabilities, see Sect.~\ref{sec:spin}.

Polarisabilities measure hence the global stiffness of the nucleon's internal
degrees of freedom against displacement in an electric or magnetic field of
definite multipolarity and non-vanishing frequency $\omega$.  They contain
detailed information about the constituents because they lead to quite
different dispersive effects: 
There are low-lying nuclear resonances like the $\Delta(1232)$, the charged
meson-cloud around the nucleon, internal relaxation effects, etc.
Spin-polarisabilities are particularly interesting as they parameterise the
response of the nucleon-\emph{spin} to the photon field, having no classical
analogon.

It must be stressed that dynamical polarisabilities are a concept
complementary to \emph{generalised} polarisabilities. The latter probe the
nucleon in virtual Compton scattering, i.e.~with an incoming photon of zero
energy and non-zero virtuality, and can provide information about the spatial
distribution of charge and magnetism
inside the nucleon. 
\emph{Dynamical polarisabilities} on the other hand test the global response
of the internal nucleonic degrees of freedom to a \emph{real} photon of
\emph{non-zero} energy and answer the question \emph{which} internal degrees
of freedom govern the structure of the nucleon at low energies by
parameterising the \emph{time-scale} on which the interaction takes place.

Nucleon Compton scattering provides thus a wealth of information about the
internal structure of the nucleon. However, in contradistinction to many other
electro-magnetic processes, 
the nucleon-structure effects probed in Compton scattering have not been
analysed in terms of a multipole-expansion at fixed energies.  Instead, most
experiments have focused on just two parameters, namely the static electric
and magnetic polarisabilities $\bar{\alpha}:=\alpha_{E1}(\omega=0)$ and
$\bar{\beta}:=\beta_{M1}(\omega=0)$, which are also often for brevity called
``the polarisabilities'' of the nucleon.  Therefore, quite different
theoretical frameworks are at present able to provide a consistent,
qualitative picture for the leading static polarisabilities. Their dynamical
origin is however only properly revealed by their energy-dependence, which
varies with the underlying mechanism. For the proton, the generally accepted
static values\footnote{It is customary to measure the scalar
  dipole-polarisabilities in $10^{-4}\;\fm^3$, so that the units are dropped
  in the following. Notice that the nucleon is quite stiff.}  are
$\bar{\alpha}^p\approx 12\times 10^{-4}\;\fm^3,\;\bar{\beta}^p\approx 2\times
10^{-4}\;\fm^3$, with error-bars of about $1\times 10^{-4}\;\fm^3$.  For the
neutron, different types of experiments report a range of values
$\bar{\alpha}^n\in[-4;19]$, and even less is known about the
spin-polarisabilities.

These notes give an overview how combining recent theoretical advances with
data from the new generation of high-precision Compton scattering experiments
also with polarised beams and targets can offer fresh, detailed insight into
these problems: We define dynamical polarisabilities and study their
low-energy contents in Sect.~\ref{sec:def}, together with a model-independent
extraction of the scalar dipole-polarisabilities of the proton.
Section~\ref{sec:spin} proposes to extract their energy-dependence and
simultaneously determine the ill-known spin-polarisabilities by polarised
experiments. Finally, an accurate determination of the neutron
polarisabilities from deuteron Compton scattering is reported in
Sect.~\ref{sec:deuteron}.

\section{Definition and Low-Energy Contents}
\label{sec:def}

A rigorous definition of energy-dependent or dynamical polarisabilities starts
instead of (\ref{polsfromints}) from the six independent amplitudes into which
the $T$-matrix of real Compton scattering is decomposed:
\begin{eqnarray}
  T(\omega,z)&=& A_1(\omega,z)\;(\vec{\epsilon}\,'^*\cdot \vec{\epsilon}) +
  A_2(\omega,z)\;(\vec{\epsilon}\,'^*\cdot\hat{\vec{k}})\;(\vec{\epsilon}
  \cdot\hat{\vec{k}}')
  +\ii\,A_3(\omega,z)\;\vec{\sigma}\cdot
  \left(\vec{\epsilon}\,'^*\times\vec{\epsilon}\,\right)\non
  \\&&
\label{Tmatrix}
  +\ii\,A_4(\omega,z)\;\vec{\sigma}\cdot
  \left(\hat{\vec{k}}'\times\hat{\vec{k}}\right)
  (\vec{\epsilon}\,'^*\cdot\vec{\epsilon})
  +\ii\,A_5(\omega,z)\;\vec{\sigma}\cdot
  \left[\left(\vec{\epsilon}\,'^*\times\hat{\vec{k}} \right)\,
    (\vec{\epsilon}\cdot\hat{\vec{k}}') -\left(\vec{\epsilon}
      \times\hat{\vec{k}}'\right)\,
    (\vec{\epsilon}\,'^*\cdot\hat{\vec{k}})\right]
  \\&&
  +\ii\,A_6(\omega,z)\;\vec{\sigma}\cdot
  \left[\left(\vec{\epsilon}\,'^*\times\hat{\vec{k}}'\right)\,
    (\vec{\epsilon}\cdot\hat{\vec{k}}') -\left(\vec{\epsilon}
      \times\hat{\vec{k}} \right)\,
    (\vec{\epsilon}\,'^*\cdot\hat{\vec{k}})\right]\non
\end{eqnarray}
Here, $\hat{\vec{k}}$ ($\hat{\vec{k}}'$) is the unit-vector in the momentum
direction of the incoming (outgoing) photon with polarisation $\vec{\epsilon}$
($\vec{\epsilon}\,'^*$) and $\theta$ the scattering angle, $z=\cos\theta$.

We separate these amplitudes into a pole-part and a non-pole or structure-part
$\bar{A}_i$. Intuitively, one could define the pole-part as the one which
leads to the Powell cross-section of a point-like nucleon with anomalous
magnetic moment and thus parameterises all we hope to have understood about
the nucleon. Then, it would seem, the structure-part contains all information
about the internal degrees of freedom which make the nucleon an extended,
polarisable object. However, the question which part a contribution belongs to
cannot be answered uniquely. In the following, only those terms which have a
pole either in the $s$-, $u$- or $t$-channel are treated as non-structure, see
Ref.~\cite{polas2} for details. This means for example that the large
contribution to the spin-polarisabilities from the $\pi^0\gamma\gamma$-vertex
via a $\pi^0$-pole in the $t$-channel is not part of the polarisabilities thus
defined. In the calculation of observables, such a separation is clearly
irrelevant because the structure-dependent and structure-independent part must
be summed. Here, however, we investigate the r\^ole of the \emph{internal}
nucleonic degrees of freedom for the polarisabilities. They are contained only
in the structure-part of the amplitudes.

We also choose to work in the centre-of-mass frame. Thus, $\omega$ denotes the
cm-energy of the photon, $M$ the nucleon mass, and $W=\sqrt{s}$ the total
cm-energy. Following older work on the multipole-decomposition of the Compton
amplitudes and pulling a kinematical factor out relative to
(\ref{polsfromints}), one obtains for the expansion of the
\emph{structure-parts} of the amplitudes in terms of polarisabilities
\begin{align}
  \bar{A}_1(\omega,\,z) &\dis=\frac{4\pi\,W}{M}\,\left[\alpha_{E1}(\omega)
    +z\,\beta_ {M1}(\omega)\right]\,\omega^2+\dots
  , \nonumber \;\;\;\; \bar{A}_2(\omega,\,z) \dis
  =-\frac{4\pi\,W}{M}\,\beta_{M1}(\omega)\,\omega^2
  +\dots
  , \nonumber
  \\[1ex]
  \bar{A}_3(\omega,\,z) &\dis=-\frac{4\pi\,W}{M}\,\left[\gamma_{E1E1}(\omega)
    +z\,\gamma_{M1M1}(\omega)+\gamma_{E1M2}(\omega)
    +z\,\gamma_{M1E2}(\omega)\right]\,\omega^3+\dots
  , \nonumber
  \\[1ex]
  \bar{A}_4(\omega,\,z) &\dis=\frac{4\pi\, W}{M}\,\left[-\gamma_{M1M1}(\omega)
    +\gamma_{M1E2}(\omega)\right]\,\omega^3+\dots
  ,
\label{eq:strucamp}
\\[1ex]
\bar{A}_5(\omega,\,z) &\dis=\frac{4\pi\,
  W}{M}\,\gamma_{M1M1}(\omega)\,\omega^3
  +\dots
  , \nonumber
  \;\;\;\; \bar{A}_6(\omega,\,z) =\frac{4\pi\,
    W}{M}\,\gamma_{E1M2}(\omega)\,\omega^3+\dots
\end{align}
The various polarisabilities are thus identified \emph{at fixed energy} only
by their different angular dependence. Clearly, the complete set of dynamical
polarisabilities does -- like all quantities defined by
multipole-decompositions -- not contain more or less information about the
temporal response or dispersive effects of the nucleonic degrees of freedom
than the un-truncated Compton amplitudes. However, the information is more
readily accessible and easier to interpret: 
Each mechanism and channel leaves a characteristic signature in the
polarisabilities.  Moreover, it will turn out that all polarisabilities beyond
the dipole ones can be dropped in (\ref{eq:strucamp}), as they are so far
invisible in observables. For that reason, they were sacrificed to brevity in
the expressions above and purists should consider Ref.~\cite{polas2}.

\begin{figure}[!htb]
\begin{center}
  \includegraphics*[width=15 cm]{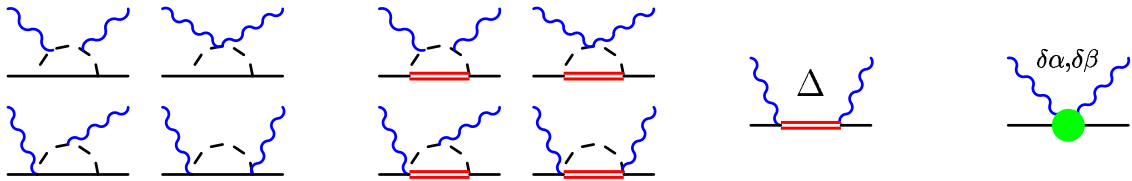}
\begin{minipage}[t]{16.5 cm}
\caption{\label{fig:polas}The dominant interactions in $\chi$EFT which
  give rise to the nucleon polarisabilities. Left to right: pion cloud around
  the nucleon and $\Delta$; $\Delta$ excitations; short-distance effects.
  Permutations and crossed diagrams not shown.  From Ref.~\cite{polas2}.}
\end{minipage}
\end{center}
\end{figure}

\absatz To identify the microscopically dominant low-energy degrees of freedom
inside the nucleon in a model-independent way, we employ the unique low-energy
theory of QCD, namely Chiral Effective Field Theory ($\chi$EFT). This
extension of Chiral Perturbation Theory to the few-nucleon sector contains
only those low-energy degrees of freedom which are observed at the typical
energy of the process, interacting in all ways allowed by the underlying
symmetries of QCD. A power-counting allows for results of finite,
systematically improvable accuracy and thus for an error-estimate.  The
resulting contributions at leading order, listed in Fig.~\ref{fig:polas}, are
easily motivated:

\newcounter{mycounti}
\begin{list}{(\arabic{mycounti})}{\usecounter{mycounti}
    \setlength{\itemsep}{0.5ex} \setlength{\topsep}{0.5ex}
    \setlength{\parsep}{0em}}
\item Photons couple to the charged pion cloud around the nucleon and around
  the $\Delta$, signalled by a characteristic cusp at the one-pion production
  threshold.
  
\item It is well-known that the $\Delta(1232)$ as the lowest nuclear resonance
  can be excited in the intermediate state by the strong $\gamma N\Delta$
  $M1$-transition, leading to a para-magnetic contribution to the static
  magnetic dipole-polarisability $\bar{\beta}_{\Delta} =+[7\dots13]$. Its
  signal is a characteristic resonance-shape, as in the Lorentz model of
  polarisabilities in classical electro-dynamics.
  
\item As the observed static value $\bar{\beta}^p\approx 2$ is smaller by a
  factor of $5$ than the $\Delta$ contribution, a strongly dia-magnetic
  component must exist. The resultant fine-tuning at zero photon-energy is
  unlikely to hold once the evolution of the polarisabilities with the photon
  energy is considered: If dia- and para-magnetism are of different origin, it
  is more than likely that they involve different scales and hence different
  energy-dependences.  Therefore, they are apt to be dis-entangled by
  \emph{dynamical polarisabilities}. We sub-sume this short-distance Physics
  which is at this order not generated by the pion or $\Delta$ into two
  \emph{energy-independent} low-energy coefficients
  $\delta\alpha,\;\delta\beta$. While na\"ive dimensional analysis sees them
  suppressed by an order of magnitude,
  $|\delta\alpha|\sim|\delta\beta|\approx\frac{\alpha}{\Lambda_\chi^2M}\sim1$
  with $\Lambda_\chi\approx 1$ GeV the breakdown-scale of \ChiEFT, the facts
  require them to be included at leading order already.
\end{list}

\vspace*{-3ex}

\begin{figure}[!htb]
\begin{center}
  \includegraphics[width=0.87\linewidth]{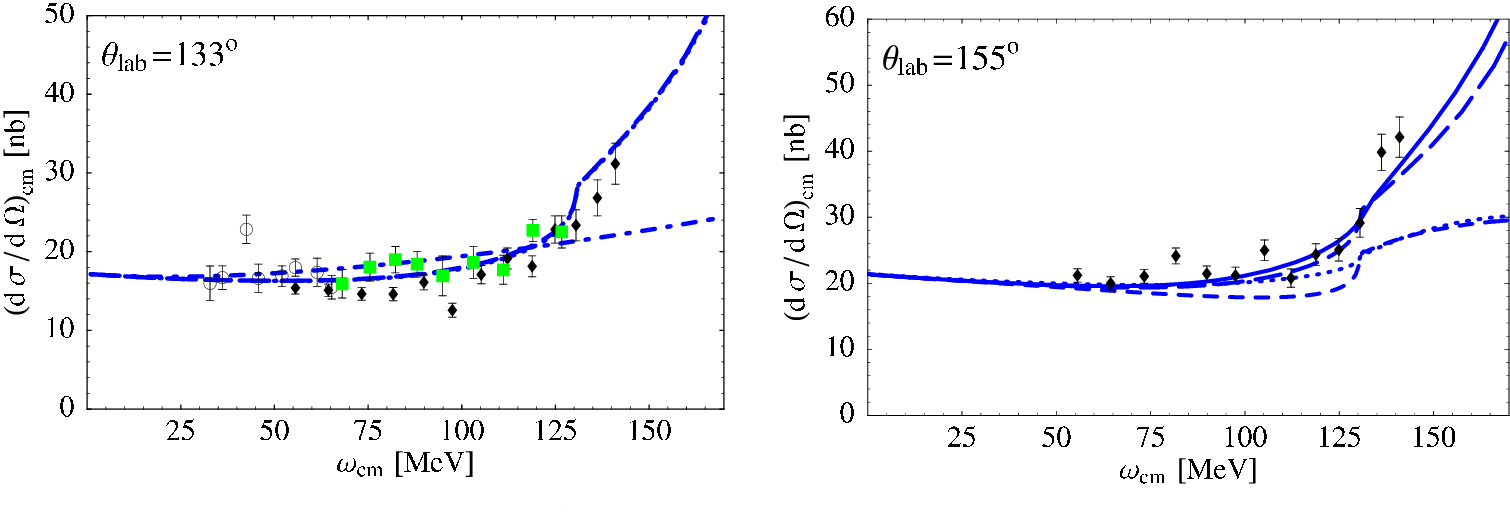}
\begin{minipage}[t]{16.5 cm}
\caption{\label{fig:protondata}
  Typical differential cross-sections for proton Compton scattering. Left:
  Data and $\chi$EFT without polarisabilities (dash-dotted); with only
  dipole-polarisabilities (dashed); full amplitude (dotted). Right: Dispersion
  Theory (solid) compared to $\chi$EFT with (long dashed) and without $\Delta$
  contributions ($\calO(p^3)$: short dashed; $\calO(p^4)$: dotted).  From
  Refs.~\cite{polas2,Beane:2004ra}.}
\end{minipage}
\end{center}
\end{figure}

The free constants $\delta\alpha$ and $\delta\beta$ are determined by fitting
the un-expanded $\chi$EFT-amplitude to the cornucopia of Compton scattering
data on the proton~\cite{olmosdeleon,pdata} below $200\;\MeV$,
cf.~Fig.~\ref{fig:protondata}. If these values are consistent within
error-bars with the Baldin sum-rule for the proton,
$\bar{\alpha}^p+\bar{\beta}^p=13.8\pm0.4$, one can in a second step use this
number as additional input. One obtains indeed
\begin{equation}
  \begin{array}{rll}
  \label{protonvalues1}
  \mbox{free fit:}&
  \dis\bar{\alpha}^p=11.5\pm2.4_\mathrm{stat}&
  ,\;\;\;\;\bar{\beta}^p=3.4\pm1.7_\mathrm{stat}\;\;\\
  \mbox{with Baldin:} &
  \dis\bar{\alpha}^p=11.0\pm1.4_\mathrm{stat}\pm0.4_\mathrm{Baldin}&
  ,\;\;\;\;\bar{\beta}^p=2.8\mp1.4_\mathrm{stat}\pm0.4_\mathrm{Baldin}
  \end{array}
\end{equation}
%
Higher-order corrections are estimated to contribute an error of about $\pm1$
not displayed here. The statistical error dominates. These results compare
both in magnitude and uncertainty favourably with state-of-the-art results
e.g.~from a global data-analysis~\cite{olmosdeleon} or from Dispersion
Theory~\cite{Report}:
\begin{equation}
  \begin{array}{rll}
  \mbox{dispersion theory:}&
  \bar{\alpha}^p=12.4\pm0.6_\mathrm{stat}\mp0.6_\mathrm{syst}&
  ,\;\;\;\;\bar{\beta}^p=1.4\pm0.7_\mathrm{stat}\pm0.5_\mathrm{syst}
\end{array}
\end{equation}
The two short-distance parameters are indeed anomalously large,
$\delta\alpha=-5.9\pm1.4,\; \delta\beta=-10.7\pm1.2$, justifying their
inclusion at leading order. As expected, $\delta\beta$ is dia-magnetic.
Truncating the multipole-expansion in (\ref{eq:strucamp}) is justified because
the influence of the quadrupole and higher polarisabilities on cross-sections
and asymmetries for energies up to about $300\;\MeV$ is hardly visible,
cf.~Figs.~\ref{fig:protondata} and \ref{fig:asyms}.

Not surprisingly, the $\Delta$-contribution is most pronounced at large
momentum-transfers, i.e.~backward angles, where even the next-to-leading order
calculation without dynamical $\Delta$ cannot reproduce the steep rise seen in
the data as low as $120\;\MeV$. For that reason, about half of the proton data
below $200\;\MeV$ were excluded in the analysis leading to the final numbers
in \cite{Beane:2004ra}.

\absatz With the parameters now fixed, the energy-dependence of all
polarisabilities is predicted.  The proton and neutron polarisabilities are
very similar in $\chi$EFT, iso-vectorial effects being of higher order. This
point will be confirmed also in Sect.~\ref{sec:deuteron}. We compare with a
result from Dispersion Theory, in which the dispersive effects are sub-sumed
into integrals over experimental input from a different kinematical r\'egime,
namely the photo-absorption cross-section $\gamma N\to X$. Its major
error-sources are the insufficient neutron data and the uncertainty in
modelling the high-energy behaviour of the dispersive integral.

\begin{figure}[!htb]
\begin{center}
  \includegraphics[width=0.93\linewidth]{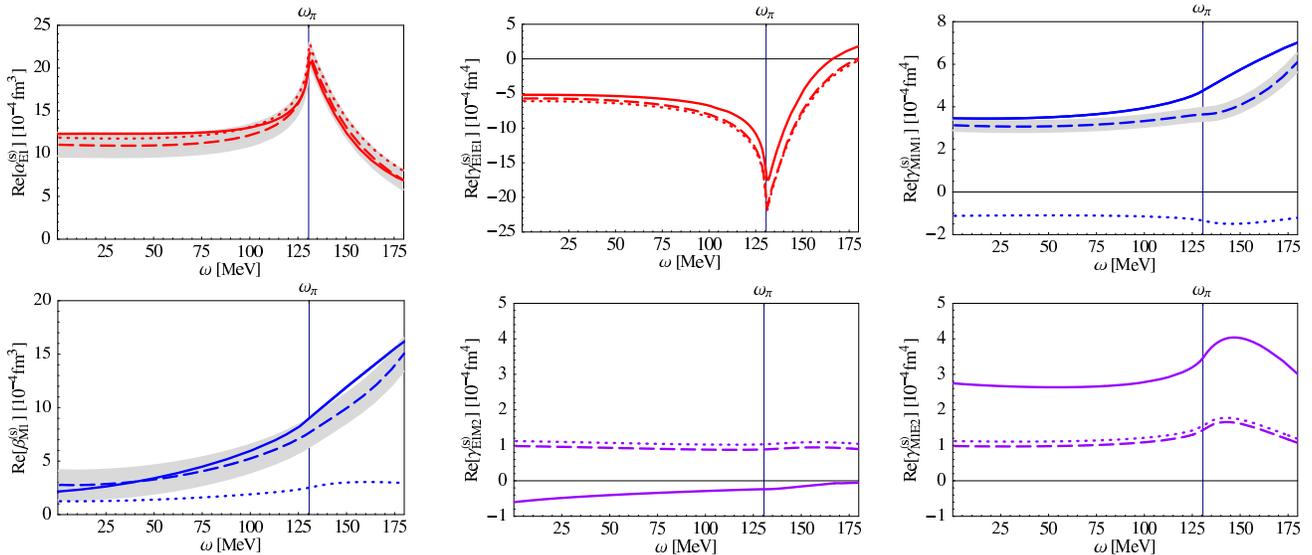}
\begin{minipage}[t]{16.5 cm}
\caption{\label{fig:dipolepols} The dipole-polarisabilities, predicted by
  Dispersion Theory (solid) and $\chi$EFT with (dashed, band from fit-errors)
  and without (dotted) explicit $\Delta$. Left: spin-independent; middle,
  right: spin-dependent. $\omega_\pi$: one-pion production threshold. From
  Ref.~\cite{polas2}.}
\end{minipage}
\end{center}
\end{figure}

The dipole-polarisabilities show the expected behaviour, and thus no
low-energy degrees of freedom inside the nucleon are missing.
Figure~\ref{fig:dipolepols} shows that dynamical effects are large at photon
energies of $80- 200\;\mathrm{MeV}$ where most experiments to determine
polarisabilities are performed. Especially at large backward angle,
unpolarised and polarised cross-sections are rather sensitive to the
non-analytical structure of the amplitude around the pion cusp and
$\Delta$-resonance, see Figs.~\ref{fig:protondata} and \ref{fig:asyms},
and~\cite{polas2,polas3,dpolas}.

The pion-cusp -- clearly seen in the $E1$-polarisabilities -- is
quantitatively understood already at leading order. The dipole
spin-polarisabilities are predictions, three of them being independent of the
parameter-determination. Since the mixed spin-polarisabilities (lower centre
and right panel of Fig.~\ref{fig:dipolepols}) are small, the relative
uncertainties in both Dispersion Theory and $\chi$EFT are large.  While the
static polarisabilities are real, the dynamical polarisabilities become
complex once the energy in the intermediate state is high enough to create an
on-shell intermediate state, the first being the physical $\pi N$-continuum,
see~\cite{polas2}.

Most notably is however the strong energy-dependence induced into
$\beta_{M1}(\omega)$ and all polarisabilities containing an $M1$ photon even
well below the pion-production threshold by the unique signature of the
$\Delta$ resonance: At $\omega\approx 90$ MeV, $\beta_{M1}$ is already about
$3$ units larger than its static value, rendering the traditional
approximation of $\beta_{M1}(\omega)$ as ``static-plus-small-slope''
$\bar{\beta}+\omega^2\bar{\beta}_\nu$ inadequate. This also reveals the good
quantitative agreement between the measured value of $\bar{\beta}^p$ and the
prediction in a $\chi$EFT without explicit $\Delta$ as accidental: The
contribution from the pion-cloud alone is not dispersive enough to explain the
energy-dependence of $\beta_{M1}$.  One could include higher-order terms in
the photon-nucleon interactions which mimic the $\Delta$, but their strengthes
are then a-priori un-determined. Furthermore, as the effect is strong, this
would in-validate the power-counting at the heart of a model-independent
analysis. When the $\Delta$ is included, $\gamma_{M1M1}$ even changes sign and
$\gamma_{M1E2}$ nearly triples in magnitude. While the fine details of the
rising para-magnetism differ between $\chi$EFT and Dispersion Theory, they are
consistent within the uncertainties of the $\chi$EFT curve. The discrepancy
between the two schemes above the one-pion production threshold is connected
to a derailed treatment of the width of the $\Delta$-resonance, which is
neglected in leading-one-loop $\chi$EFT at low energies.

The two short-distance parameters $\delta\alpha,\;\delta\beta$ which sub-sume
all Physics not generated by the pion cloud or the $\Delta$ suffice to
describe the polarisabilities up to energies of $300\;\MeV$ when the finite
width of the $\Delta$ is included~\cite{polas2}. Therefore, three constraints
arise on any attempt to explain them microscopically:

\newcounter{mycountii}
\begin{list}{(\arabic{mycountii})}{\usecounter{mycountii}
    \setlength{\itemsep}{0.5ex} \setlength{\topsep}{0.5ex}
    \setlength{\parsep}{0em}}
\item The effect must be $\omega$-independent over a wide range, like
  $\delta\alpha,\;\delta\beta$.
  
\item It must occur in the electric and magnetic scalar polarisabilities,
  leading to the values for $\delta\alpha,\;\delta\beta$ predicted in
  $\chi$EFT, but it must be absent at least in the pure spin-polarisabilities
  $\gamma_{E1E1},\;\gamma_{M1M1}$.
  
\item Its prediction for the proton and neutron must be similar because
  iso-vectorial effects were shown to be small and
  energy-independent~\cite{polas2,dpolas}, see also the neutron
  polarisabilities discussed in Sect.~\ref{sec:deuteron}.
\end{list}

\noindent
Two proposals to explain $\delta\alpha,\;\delta\beta$ were put forward: One
attributes them to an interplay between short-distance Physics and the pion
cloud occuring from the next-to-leading order chiral Lagrangean~\cite{BKMS};
the other to the $t$-channel exchange of a meson or correlated two-pion
exchange~\cite{Guichon}. Whether either gives a 
quantitative description of the short-distance coefficients meeting these
criteria is not clear yet.

\section{Spin-Polarisabilities and Energy-Dependence from Data}
\label{sec:spin}

Future doubly-polarised, high-accuracy experiments in particular around the
pion-production threshold provide an exciting avenue to extract the
energy-dependence of the six polarisabilities per nucleon, both
spin-independent~\cite{polas2} and spin-dependent~\cite{polas3}.  This is
particularly interesting since practically no direct information exists at
present on the spin-polarisabilities:
Only two linear combinations are constrained from experiments~\cite{Report},
and only at zero photon-energy. These forward and backward
spin-polarisabilities $\bar{\gamma}_0$ and $\bar{\gamma}_\pi$ of the nucleon
involve however all four static (dipole) spin-polarisabilities.

\begin{figure}[!htb]
\begin{center}
  \includegraphics[width=0.85\linewidth]{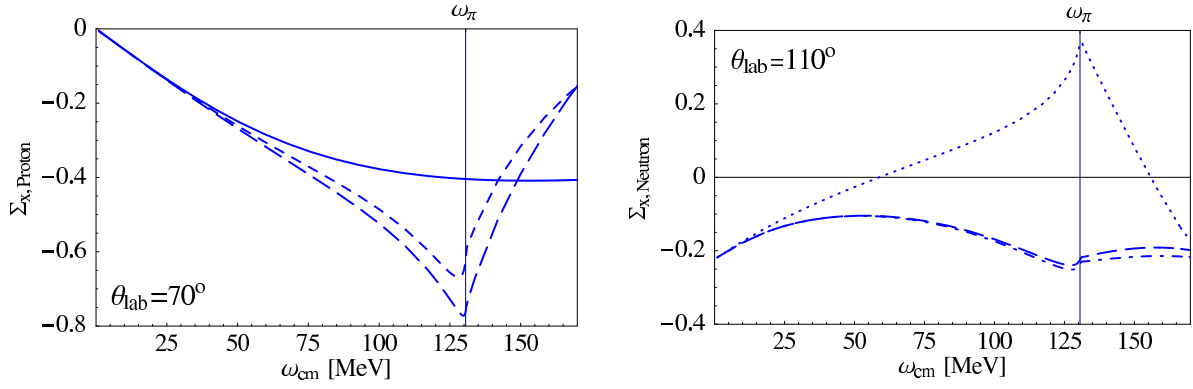}
\begin{minipage}[t]{16.5 cm}
\caption{
  \label{fig:asyms} Typical sensitivity of the proton
  (left) and neutron (right) Compton scattering asymmetry $\Sigma_x$.
  Left, solid lines: no polarisabilities; short dashed: no $\Delta$ Physics;
  long dashed: full $\chi$EFT. Right, dashed lines: full $\chi$EFT result;
  dotted: without spin-polarisabilities; dot-dashed: without quadrupole
  polarisabilities. From Ref.~\cite{polas3}.}
\end{minipage}
\end{center}
\end{figure}

Consider the Compton scattering asymmetry $\Sigma_x$: The nucleon-spin lies in
the reaction-plane and perpendicular to the circularly polarised incoming
photon. Figure~\ref{fig:asyms} shows strong sensitivity on the
spin-polarisabilities $\gamma_i(\omega)$ and on $\Delta$-Physics, while higher
polarisabilities are negligible. Similar findings hold for other asymmetries
also with linearly polarised photons~\cite{polas3,tocome}.

With 
higher polarisabilities negligible, one can use the multipole-expansion of the
scattering amplitudes (\ref{eq:strucamp}) to perform with increasing
sophistication fits of the six dipole-polarisabilities per nucleon to
data-sets which combine polarised and spin-averaged experiments, taken at
fixed energy but varying scattering angle~\cite{polas3}. As starting values
for the fit, one might assume that the energy-dependence of the
polarisabilities derived above in $\chi$EFT is correct, with deviations taken
as energy-independent. The corresponding free normalisation for each
dipole-polarisability can be used to determine the static values. At low
energies, this should be a viable procedure because only $\Delta$- and
pion-degrees of freedom are expected to give dispersive contributions to the
polarisabilities, and \ChiEFT predicts their behaviour model-independently.
When the fit of eq.~(\ref{eq:strucamp}) to data is made at each energy
independently,
repeating this procedure for various energies gives the energy-dependence of
the polarisabilities. In this way, one can extract dynamical polarisabilities
directly from the angular dependence of observables.

The spin-independent polarisabilities $\alpha_{E1}(\omega)$,
$\beta_{M1}(\omega)$ from $\chi$EFT in Fig.~\ref{fig:dipolepols} agree very
well with Dispersion Theory, both in their energy-dependence and overall size.
They could therefore be used in a second step as input to reduce the number of
fit functions in (\ref{eq:strucamp}) to four, namely the four dipole
spin-polarisabilities. The good agreement in $\gamma_{E1E1}(\omega)$ and
possibly $\gamma_{M1M1}(\omega)$ can -- similarly -- be used to reduce the
number of fit functions further to three or two per nucleon:
$\gamma_{E1M2}(\omega)$ and $\gamma_{M1E2}(\omega)$.

\begin{figure}[!htb]
\begin{center}
  \includegraphics*[width=0.95\linewidth]{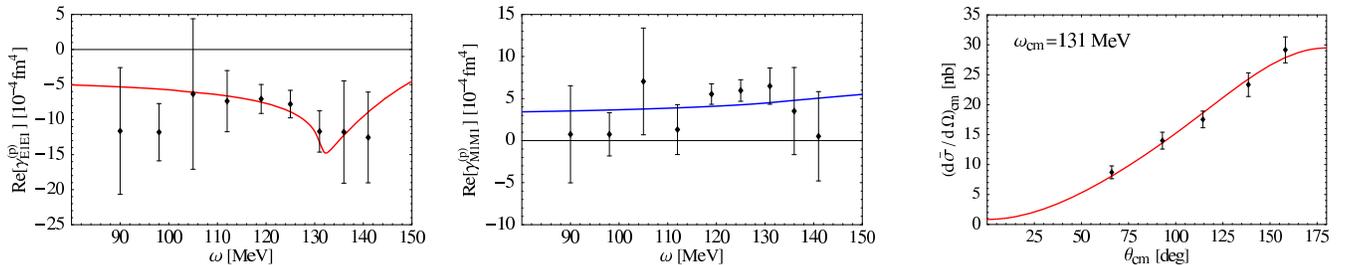}
\begin{minipage}[t]{16.5 cm}
\caption{\label{fig:feasibility}Feasibility study of a multipole-decomposition
  using existing data. Statistical error-bars only; solid: \ChiEFT-prediction.
  From Ref.~\cite{tocome}.}
\end{minipage}
\end{center}
\end{figure}

At present, only un-polarised data are available, in which the dipole
spin-polarisabilities are however anything but negligible. As a
feasibility-study, we demonstrate the method by a superficial fit of the pure
spin-polarisabilities to the existing data under the assumption that the mixed
spin-polarisabilities are negligible and the spin-independent polarisabilities
are predicted correctly~\cite{tocome}. Clearly, today's un-polarised data are
only sensitive to a linear combination of spin-polarisabilities and the
error-bars are rather large as the dependence on the spin-polarisabilities is
quadratic in $\cos\theta$, while the data are nearly linear at fixed $\omega$,
see Fig.~\ref{fig:feasibility}.

To analyse Compton scattering via a multipole-decomposition at fixed energies
can thus substantially further our knowledge on the spin-structure of the
proton. 

\section{Neutron Polarisabilities and Deuteron Compton Data}
\label{sec:deuteron}

Does the neutron react similarly under deformations,
$\bar{\alpha}^p\approx\bar{\alpha}^n$, $\bar{\beta}^p\approx\bar{\beta}^n$,
etc, as \ChiEFT predicts? Since free neutrons can only rarely be used in
experiments, their properties are usually extracted from data taken on
few-nucleon systems by subtracting nuclear-binding effects. This should be
done in a model-independent way and with an estimate of the theoretical
uncertainties as in \ChiEFT.
However, different types of experiments report a range of values
$\bar{\alpha}^n\in[-4;19]$: Coulomb scattering of neutrons off lead, or
deuteron Compton-scattering with and without breakup, see~\cite{dpolas} for a
list. As deuteron Compton scattering 
should provide a clean way to extract the iso-scalar polarisabilities
$\bar{\alpha}^s:=\half(\bar{\alpha}^p+\bar{\alpha}^n)$ and $\bar{\beta}^s$
parallel to determinations of the proton polarisabilities, experiments
were performed~\cite{luca94} in Urbana 
at $\omega=49$ and $69$ MeV, in
Saskatoon (SAL) 
at $94$ MeV, and in Lund 
at $55$ and $66$ MeV. While the low-energy extractions are consistent with
small iso-vectorial polarisabilities, the SAL data lead to conflicting
analyses: 
The original publication 
gave $\bar{\alpha}^s=8.8\pm1.0$, employing the 
Baldin sum-rule for the static nucleon polarisabilities,
$\bar{\alpha}^s+\bar{\beta}^s=14.5\pm0.6$.
Levchuk and L'vov obtained
$\bar{\alpha}^s=11\pm2,\;\bar{\beta}^s=7\pm2$~\cite{Levchuk:2000mg}; and Beane
et al.~found recently from all data
$\bar{\alpha}^s=13\pm4,\;\bar{\beta}^s=-2\pm3$~\cite{Beane:2004ra}.
The 
extraction being very sensitive to the polarisabilities,
can embedding the neutron into a nucleus lead to the discrepancy?

\begin{figure}[!htbp]
  \begin{center}
    \includegraphics*[width=\linewidth]{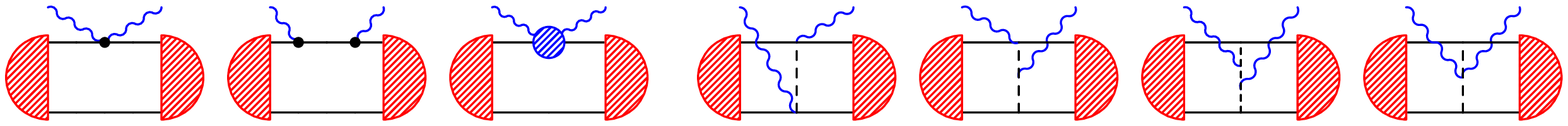}
\begin{minipage}[t]{16.5 cm}
  \caption{\label{fig:dgraphs}
    Deuteron Compton scattering in $\chi$EFT to $\calO(\epsilon^3)$. Left:
    one-body (dots: electric/magnetic couplings; blob: nucleon
    polarisabilities, Fig.~\ref{fig:polas}). Right: two-body parts
    (pion-exchange currents). Permutations and crossed diagrams not shown.
    From Ref.~\cite{dpolas}.}
\end{minipage}
\end{center}
\end{figure}

\vspace{-2.5ex}

\begin{figure}[!htb]
  \begin{center}
    \includegraphics*[width=0.82\linewidth]{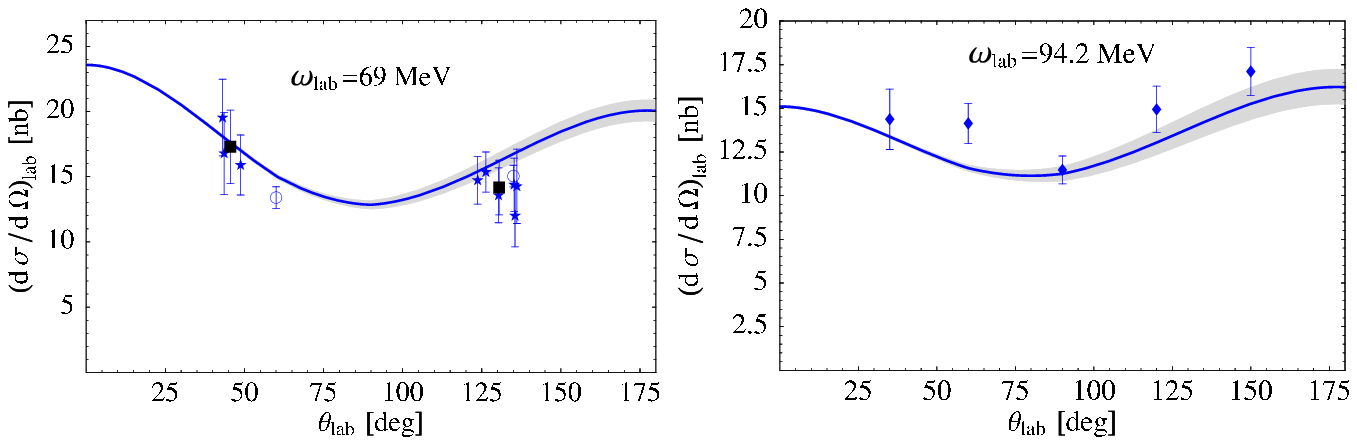}
    \\[1ex]
    \includegraphics*[width=0.82\linewidth]{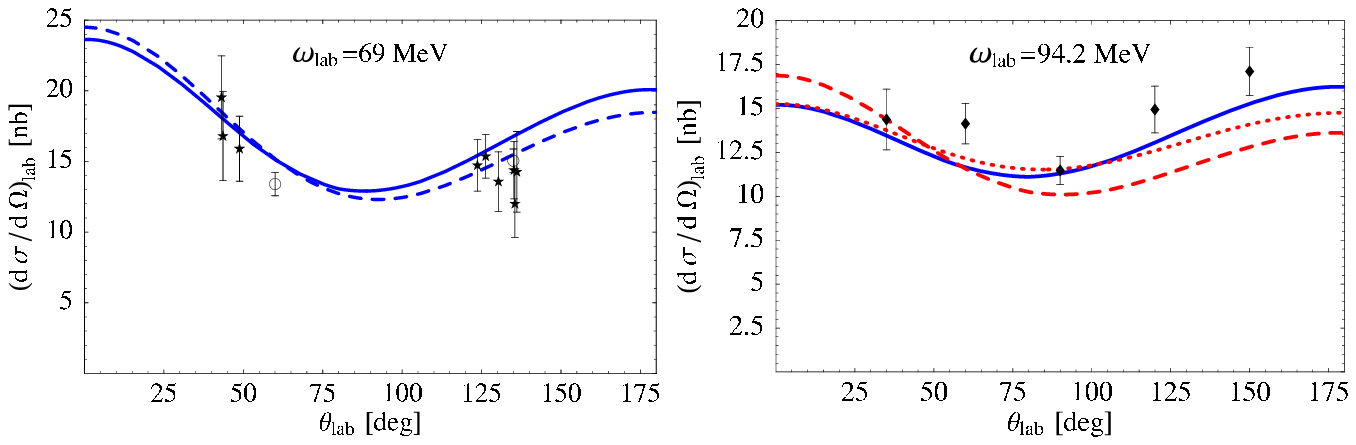}
\begin{minipage}[t]{16.5 cm}
\caption{\label{fig:dcompton} Deuteron Compton scattering in $\chi$EFT at
  $\calO(\epsilon^3)$ with $\bar{\alpha}^s,\;\bar{\beta}^s$ from
  eq.~(\ref{deuteronvalues}), using the Baldin sum rule. Top: Grey bands:
  Statistical error.
  \label{fig:dDelta}
  Bottom: Comparison between $\chi$EFT with explicit $\Delta$ (solid) and
  without explicit $\Delta$ (dashed: $\calO(p^3)$, parameter-free; dotted:
  $\calO(p^4)$, best fit). From Ref.~\cite{dpolas} with the help of
  Ref.~\cite{Beane:2004ra}.}
\end{minipage}
\end{center}
\end{figure}

Of course, two-body contributions from meson-exchange currents and
wave-function dependence must be subtracted from data with minimal theoretical
prejudice.  Figure~\ref{fig:dgraphs} lists the contributions to Compton
scattering off the deuteron to next-to-leading order in \ChiEFT. The
calculation is parameter-free, except for the short-distance coefficients
$\delta\alpha,\;\delta\beta$ in the nucleon polarisabilities. The nucleon- and
nuclear-structure contributions clearly separate at this (and the next) order.
While the two-nucleon piece does not contain the $\Delta(1232)$-resonance in
the intermediate state at this order as the deuteron is an iso-scalar target,
this is not true for the polarisabilities, as seen in Sect~\ref{sec:def}.
Figure~\ref{fig:dDelta} shows that the strong energy-dependence from exciting
the $\Delta$ is indeed pivotal to reproduce the shape of the data at $94$ MeV
in particular at back-angles.
Thus, we argue that the discrepancy in
extractions from the SAL data and 
at lower energies is resolved.

So far, dynamical effects had largely been neglected in the analyses:
Truncating the Taylor-expansion of the polarisabilities around zero-photon
energy at order $\omega^2$ under-estimates
$\beta_{M1}(\omega=95\;\MeV)-\bar{\beta}\approx 1.7$~\cite{Levchuk:2000mg},
while our multipole-analysis gives $\approx 4$ (Fig.~\ref{fig:dipolepols}).
The \ChiEFT-calculation of Beane et al.~\cite{Beane:2004ra
} cited already in the proton-extraction uses the same deuteron wave-functions
and meson-exchange currents as we, but sub-sume all $\Delta$-effects into
short-distance operators which enter only at higher order and are only weakly
dispersive. They hence exclude the two SAL-points at large angles from their
final analysis.


Fitting in $\chi$EFT with explicit $\Delta$ the two short-distance parameters
$\delta\alpha,\;\delta\beta$ to deuteron Compton scattering data above $60$
MeV (Fig.~\ref{fig:dcompton}), one finds as static values:
\begin{eqnarray}
  \begin{array}{rll}
  \label{deuteronvalues}
  \mbox{free fit:} &
  \dis\bar{\alpha}^s=12.8\pm1.4_\mathrm{stat}\pm1.1_\mathrm{wavefu}&
  ,\;\;\bar{\beta}^s=2.1\pm1.7_\mathrm{stat}\pm0.1_\mathrm{wavefu}
  \\
  \mbox{with Baldin:} &
  \dis\bar{\alpha}^s=12.6\pm0.8_\mathrm{stat}\pm0.7_\mathrm{wavefu}
  \pm0.6_\mathrm{Baldin}&
  ,\;\;\bar{\beta}^s=1.9\mp0.8_\mathrm{stat}\mp0.7_\mathrm{wavefu}
  \pm0.6_\mathrm{Baldin}
  \end{array}
\end{eqnarray}
Again, higher-order effects can be estimated to induce an additional
systematic error of $\pm1$. The Baldin sum-rule
$\bar{\alpha}^s+\bar{\beta}^s=14.5\pm0.6$ is already well-reproduced by the
unconstrained fit. Comparing with the static proton polarisabilities
(\ref{protonvalues1}) determined by the same method, the proton and neutron
polarisabilities turn indeed out identical within the statistical uncertainty.


\section{Perspectives}
\label{sec:conclusions}

Dynamical polarisabilities test the global response of the nucleon to the
electric and magnetic fields of a real photon with non-zero energy and
definite multipolarity. They answer the question which internal degrees of
freedom govern the structure of the nucleon at low energies and can be defined
by a multipole-expansion of the Compton amplitudes. While they do not contain
more or less information than the corresponding Compton scattering amplitudes,
the facts are more readily accessible and easier to interpret. Dispersive
effects in particular from the $\Delta(1232)$ are necessary to obtain accurate
extractions for the static polarisabilities of the nucleon from the available
data. Future work includes:

  
(i) The non-zero width of the $\Delta$ and higher-order effects from the
pion-cloud become crucial when probing the nucleon-response in the resonance
region.
  
(ii) As seen, a multipole-analysis from doubly-polarised, high-accuracy
experiments provides a new avenue to extract the energy-dependence of the six
dipole-polarisabilities per nucleon, both spin-independent and
spin-dependent~\cite{polas2,polas3,tocome}.  This will in particular further
our knowledge on the spin-polarisabilities, and hence on the spin-structure of
the nucleon. A (certainly incomplete) list of planned or approved experiments
at photon-energies below $300\;\MeV$ shows the concerted effort in this field:
polarised photons on polarised protons, deuterons and ${}^3$He at
TUNL/HI$\gamma$S; tagged protons at S-DALINAC; polarised photons on polarised
protons at MAMI; and deuteron targets at MAXlab. With for example only 29
(un-polarised) points for the deuteron in a small energy range of
$\omega\in[49;94]$ MeV and error-bars on the order of $15\%$, new data can
improve the situation substantially.
  
(iii) The deuteron data at $49$ and $55$ MeV~\cite{luca94} are not included in
our analysis because the $\chi$EFT-power-counting of Fig.~\ref{fig:dgraphs} is
not tailored to the low-energy end and must be modified to yield the
correct Thomson limit. This problem is partially circumvented in
Ref.~\cite{Beane:2004ra}, and a full treatment is in its finishing
stages~\cite{tocome}. On the other hand, the pion-exchange terms can be
integrated out at lower energies, and one arrives at the ``pion-less'' EFT of
QCD. Not only is this version computationally considerably less involved than
the pion-ful version $\chi$EFT; it also has the advantage that the Thomson
limit is recovered trivially. While Compton scattering becomes the less
sensitive to the polarisabilities the lower the energy, a window exists
between about $25$ and $50$ MeV where dispersive effects are negligible and
this variant can aid high-accuracy experiments e.g.~at HI$\gamma$S to extract
the static polarisabilities in a model-independent way.  Recently, Chen et
al.~\cite{Chen:2004wv} demonstrated that due to the large iso-vectorial
magnetic moment, the vector amplitudes in $d\gamma$-scattering are anomalously
enhanced in this version.  Adding to a previous calculation by Rupak and
Grie\3hammer~\cite{Griesshammer:2000mi}, they found that the data at $49$ and
$55$ MeV are in fact well in agreement with the values given above and report
$\bar{\alpha}^s=12\pm1.5,\;\bar{\beta}^s=5\pm2$.
  
Enlightening insight into the electro-magnetic structure of the nucleon has
already been gained from combining Compton scattering off nucleons and
few-nucleon systems with $\chi$EFT and energy-dependent or dynamical
polarisabilities; and a host of activities should add to it in the coming
years.

\section*{Acknowledgements}
I am grateful for the opportunity to speak and for financial support by the
DFG to attend this meeting. Foremost, I thank my collaborators --
R.P.~Hildebrandt, T.R.~Hemmert, B.~Pasquini and D.R.~Phillips -- for a lot of
fun!



\begin{thebibliography}{99}
  \itemsep -2pt
\bibitem{Griesshammer:2001uw} H.~W.~Grie{\ss}hammer and T.~R.~Hemmert:
  \PRC\textnormal{65} (2002), 045207 [nucl-th/0110006].
  
\bibitem{polas2} R.~P.~Hildebrandt, H.~W.~Grie\3hammer, T.~R.~Hemmert and
  B.~Pasquini: \EPJA\textnormal{20} (2004), 293 [nucl-th/0307070].
  
\bibitem{polas3} R.~P.~Hildebrandt, H.~W.~Grie\3hammer and T.~R.~Hemmert:
  \EPJA\textnormal{20} (2004), 329 [nucl-th/0308054].
  
\bibitem{dpolas} R.P.~Hildebrandt, H.W.~Grie\3hammer, T.R.~Hemmert and
  D.R.~Phillips: [nucl-th/0405077]. Accepted for publication in \EPJA.
  
\bibitem{olmosdeleon} Olmos de Leon et al.: \EPJA \textnormal{10} (2001) 207.
  
\bibitem{pdata} The most recent ones besides Ref.~\cite{olmosdeleon} are
  E.L.~Hallin et al., \PRC \textnormal{48} (1993) 1497; F.J.~Federspiel et
  al., \PRL \textnormal{67} (1991) 1511; B.E.~MacGibbon et al., \PRC
  \textnormal{52} (1995) 2097.
  
\bibitem{Report} D.~Drechsel, B.~Pasquini, M.~Vanderhaeghen:
  \emph{Phys.~Rep.}~\textnormal{378} (2003), 99.
  
\bibitem{Beane:2004ra} S.R.~Beane et al.: [nucl-th/0403088], and private
  communication.
  
\bibitem{BKMS} V.~Bernard et al.: \emph{Z.~Phys.}~A\textnormal{348} (1994),
  317.
  
\bibitem{Guichon} P.~A.~M.~Guichon, private communication; W.~Weise and
  B.~Pasquini: forthcoming.
  
\bibitem{tocome} R.P.~Hildebrandt, H.W.~Grie\3hammer and T.R.~Hemmert:
  forthcoming.
  
\bibitem{luca94} M.A.~Lucas: Ph.D.~thesis, Univ.~of Illinois at
  Urbana-Champaign (1994); M.~Lundin et al.: \PRL~\textnormal{90}, 192501
  (2003) [nucl-ex/0204014]; D.L.~Hornidge et al.: \PRL~\textnormal{84} (2000)
  2334 [nucl-ex/9909015].
  
  
  
\bibitem{Levchuk:2000mg} M.~I.~Levchuk and A.~I.~L'vov: \NPA\textnormal{674}
  (2000) 449 [nucl-th/9909066]; \NPA\textnormal{684} (2001) 490
  [nucl-th/0010059].
  
  
  
\bibitem{Griesshammer:2000mi} H.W.~Grie\3hammer and G.~Rupak:
  \PLB\textnormal{529} (2002) 57 [nucl-th/0012096].
  
\bibitem{Chen:2004wv} J.W.~Chen, X.D.~Ji and Y.C.~Li: [nucl-th/0408003].
  
\end{thebibliography}
\end{document}